\documentclass[twoside,twocolumn,9pt]{article}
\usepackage{extsizes}
\usepackage[super,sort&compress,comma]{natbib}
\usepackage[version=3]{mhchem}
\usepackage[left=1.5cm, right=1.5cm, top=1.785cm, bottom=2.0cm]{geometry}
\usepackage{balance}
\usepackage{widetext}
\usepackage{times,mathptmx}
\usepackage{sectsty}
\usepackage{graphicx}
\usepackage{lastpage}
\usepackage[format=plain,justification=raggedright,singlelinecheck=false,font={stretch=1.125,small,sf},labelfont=bf,labelsep=space]{caption}
\usepackage{float}
\usepackage{fancyhdr}
\usepackage{fnpos}
\usepackage[english]{babel}
\usepackage{array}
\usepackage{droidsans}
\usepackage{charter}
\usepackage[T1]{fontenc}
\usepackage[usenames,dvipsnames]{xcolor}
\usepackage{setspace}
\usepackage[compact]{titlesec}

\usepackage{epstopdf}

\definecolor{cream}{RGB}{222,217,201}

\begin{document}

\pagestyle{fancy}
\thispagestyle{plain}

\makeFNbottom
\makeatletter
\renewcommand\LARGE{\@setfontsize\LARGE{15pt}{17}}
\renewcommand\Large{\@setfontsize\Large{12pt}{14}}
\renewcommand\large{\@setfontsize\large{10pt}{12}}
\renewcommand\footnotesize{\@setfontsize\footnotesize{7pt}{10}}
\makeatother

\renewcommand{\thefootnote}{\fnsymbol{footnote}}
\renewcommand\footnoterule{\vspace*{1pt}%
\color{cream}\hrule width 3.5in height 0.4pt \color{black}\vspace*{5pt}}
\setcounter{secnumdepth}{5}

\makeatletter
\renewcommand\@biblabel[1]{#1}
\renewcommand\@makefntext[1]%
{\noindent\makebox[0pt][r]{\@thefnmark\,}#1}
\makeatother
\renewcommand{\figurename}{\small{Fig.}~}
\sectionfont{\sffamily\Large}
\subsectionfont{\normalsize}
\subsubsectionfont{\bf}
\setstretch{1.125} 
\setlength{\skip\footins}{0.8cm}
\setlength{\footnotesep}{0.25cm}
\setlength{\jot}{10pt}
\titlespacing*{\section}{0pt}{4pt}{4pt}
\titlespacing*{\subsection}{0pt}{15pt}{1pt}

\fancyfoot{}
\fancyfoot[LO,RE]{\vspace{-7.1pt}}
\fancyfoot[RO]{\footnotesize{{ This is the pre-peer reviewed, authors' version of the Nanoscale article, published in its final form at  http://dx.doi.org/10.1039/C8NR01395K ~|\thepage }}}
\fancyfoot[LE]{\footnotesize{{\thepage| This is the pre-peer reviewed, authors' version of the Nanoscale article, published in its final form at  http://dx.doi.org/10.1039/C8NR01395K}}}
\fancyhead{}
\renewcommand{\headrulewidth}{0pt}
\renewcommand{\footrulewidth}{0pt}
\setlength{\arrayrulewidth}{1pt}
\setlength{\columnsep}{6.5mm}
\setlength\bibsep{1pt}

\makeatletter
\newlength{\figrulesep}
\setlength{\figrulesep}{0.5\textfloatsep}

\newcommand{\topfigrule}{\vspace*{-1pt}%
\noindent{\color{cream}\rule[-\figrulesep]{\columnwidth}{1.5pt}} }

\newcommand{\botfigrule}{\vspace*{-2pt}%
\noindent{\color{cream}\rule[\figrulesep]{\columnwidth}{1.5pt}} }

\newcommand{\dblfigrule}{\vspace*{-1pt}%
\noindent{\color{cream}\rule[-\figrulesep]{\textwidth}{1.5pt}} }

\makeatother

\twocolumn[
  \begin{@twocolumnfalse}
\vspace{3cm}
\sffamily
\begin{tabular}{m{4.5cm} p{13.5cm} }
& {This is the pre-peer reviewed, authors' version of the Nanoscale article, published in its final form at  http://dx.doi.org/10.1039/C8NR01395K} 
\vspace{0.5cm}\\ 
 & \noindent\LARGE{\textbf{Plasmonics with two-dimensional semiconductors ``beyond graphene": from basic research to technological applications}} \\
\vspace{0.3cm} & \vspace{0.3cm} \\
  & \noindent\large{Amit Agarwal\textit{$^{a}$}, Miriam S. Vitiello\textit{$^{b}$}, Leonardo Viti\textit{$^{b}$}, Anna Cupolillo\textit{$^{c}$}, and Antonio Politano\textit{$^{d}$}} \\
 & 
\noindent\normalsize{In this minireview, we explore the main features and the prospect of plasmonics with two-dimensional semiconductors. Plasmonic modes in each class of van der Waals semiconductors have their own peculiarities, along with potential technological capabilities. Plasmons of transition-metal dichalcogenides share features typical of graphene, due to their honeycomb structure, but with damping processes dominated by intraband rather than interband transitions, unlike graphene. Spin-orbit coupling strongly affects the plasmonic spectrum of buckled honeycomb lattices (silicene and germanene), while the anisotropic lattice of phosphorene determines different propagation of plasmons along the armchair and zigzag direction. We also review existing applications of plasmonics with two-dimensional materials in the fields of thermoplasmonics, biosensing, and plasma-wave Terahertz detection. Finally, we consider the capabilities of van der Waals heterostructures for innovative low-loss plasmonic devices.} \\
\end{tabular}

 \end{@twocolumnfalse} \vspace{0.6cm}
  ]

\renewcommand*\rmdefault{bch}\normalfont\upshape
\rmfamily
\section*{}
\vspace{-1cm}


\footnotetext{\textit{$^{a}$~Department of Physics, Indian
Institute of Technology Kanpur, 208016, Kanpur, India; E-mail:
amitag@iitk.ac.in}}

\footnotetext{\textit{$^{b}$~NEST-Istituto Nanoscienze and Scuola
Normale Superiore, Piazza San Silvestro 12, 56127 Pisa, Italy}}

\footnotetext{\textit{$^{c}$~Department of Physics, University of
Calabria, via ponte Bucci, cubo 31/C 87036, Rende (CS) Italy}}

\footnotetext{\textit{$^{d}$~Istituto Italiano di
Tecnologia-Graphene Labs via Morego 30, 16163 Genova, Italy;
E-mail: antonio.politano@iit.it}}

%



\vspace{1cm}

\section{Introduction}

After the groundbreaking impact of graphene \cite{Bonaccorso1246501,C4NR01600A}, the scientific community is actively exploring
other two-dimensional (2D) semiconductors ``beyond graphene" for
their promising applications capabilities, often complementary to
those of graphene. \cite{R1} Different classes of 2D
semiconductors have emerged in recent years: transition-metal
dichalcogenides \cite{R2}; black phosphorus \cite{R3};
silicene/germanene \cite{R4}; and IV-VI semiconductors. \cite{R5}

Many innovative applications, widely used in our daily lives, are
based on the exploitation of collective properties of matter
(ferromagnetism, superconductivity, the quantum Hall effect and
plasmonic excitations). Therefore, the comprehension of collective
electronic excitations is crucial in order to develop new
disruptive technologies for health, telecommunications, energy
etc. In particular, the novel field of plasmonics has recently
emerged, in consideration of the progress of nanotechnology and
nanofabrication. Plasmonics deals with the generation, propagation
and detection of plasmonic excitations, which are collective
electronic excitations produced by an electromagnetic field. 
\cite{Dragoman} The field confinement and enhancement resulting
from the interaction between matter and radiation can be
technologically used for devising plasmonic devices for diverse
applications, ranging from optics, biology, nanoelectronics and
nanophotonics. Herein, we review the peculiarities, the
applications, the pitfalls and the grand challenges of plasmonics
with 2D semiconductors.

\section{Peculiarities of plasmons in the Flatland}

Plasmon modes are collective charge density excitations
(oscillations), typically occurring in charged electron gases in
solids in the presence of Coulomb interactions.

In the basic `classical' picture, the long-wavelength plasmon
dispersion (for $q\ll k_F$, with $q$ the momentum and $k_F$ the
Fermi wave-vector) can be obtained from a macroscopic hydrodynamic
model. Equating the force (due to Coulomb interactions) on the
deviation of the electron density at a given location to the rate
change of the momentum of the density deviations, and by using the
continuity equation, the plasmon dispersion $\omega_{\rm pl}(q)$
in the long-wavelength limit can be obtained to
be\cite{giuliani2005quantum,R60}
\begin{equation} \label{eq1}
\omega_{\rm pl}(q) = \sqrt{ \frac{n}{m^*}~q^2~V_q}~,
\end{equation}
where $V_q$ is the Fourier transform of the Coulomb potential, $n$
is the electron density and $m^*$ is the effective electron mass.
Using the Fourier transform of the
long-range Coulomb interaction in two dimensions $V_q = 2 \pi e^2/\kappa q$, with $\kappa$ being the static dielectric constant, immediately
yields $\omega_{\rm pl} \propto \sqrt{q}$. This square-root
dependence of the plasmon dispersion on $q$ in the long-wavelength
limit is a universal feature of two-dimensional electron gases
with unscreened Coulomb interaction.

In a quantum mechanical picture, plasmons arise from the poles of
the interacting density-density response function, or alternately, 
as poles of the loss function $E_{\rm loss} = {\rm
Im}[-1/\epsilon(\omega,q)]$, where $\epsilon(\omega,q)$ is the
dynamical dielectric function. \cite{giuliani2005quantum} The
electron energy loss function can be experimentally probed via
high-resolution electron energy loss spectroscopy.\cite{Ref5} An 
effective theoretical approach to calculate the dynamical
dielectric function is the random phase approximation (RPA),
whereby only Hartree terms (connected single loop diagrams) in the
interacting response are retained. Within RPA, $\epsilon(\omega,q)
\approx 1 - V_q \Pi^{0}(q,\omega)$, where $V_q$ is the Fourier
transform of the Coulomb interaction, and $\Pi^{0}(q,\omega)$ is
the non-interacting density-density response function, which can
be calculated using either the effective low-energy, or
tight-binding, or ab-initio density functional theory (DFT)-based
Hamiltonian.\cite{PhysRevB.96.035422} Typically, we have $V_q
\propto 1/q$ for unscreened Coulomb repulsion in 2D, and in the
long-wavelength limit we have $\Pi^{0} \propto q^2/\omega^2$,
generally for all materials.\cite{thakur2} Consequently, the
zeros of the dynamical $\epsilon(\omega,q)$ also lead to 
$\omega_{\rm pl} \propto \sqrt{q}$ in the long-wavelength limit of the quantum mechanical
picture as well,  consistent with the
classical hydrodynamic description for homogeneous response.

While the $\sqrt{q}$ dependence of the plasmon dispersion in 2D
materials is ubiquitous (for unscreened Coulomb repulsion), the
dependence of the plasmon energies $\omega_{\rm pl}$  on the
electron density $n$ in a 2D semiconductor varies from material to
material. For example, $\omega_{\rm pl} (q \to 0) \propto n^{1/2}$
in a parabolically dispersing 2D electron gas (2DEG) with $E_{\bf
k} = \hbar^2 k^2/(2m)$ and $\hbar k$ being the crystal momentum, as per Eq.~\eqref{eq1}. However, for
materials with Dirac quasiparticles there is no concept of band
mass, and the hydrodynamic description breaks down. An intuitive
way to restore the `classical' hydrodynamic description for
plasmons of Dirac fermions is to ask what is the analogue of the
`inertial' mass in Dirac systems.\cite{thakur2} It turns out 
that the inertial mass in Dirac materials can be interpreted as the
cyclotron mass, which is identical to the band mass for systems
with parabolic dispersion. However, in
systems with Dirac dispersion, the inertial or the cyclotron mass $m_c$ has a 
quantum mechanical origin, and depends on the chemical
potential $\mu$ by the relation, $m_c = \mu/v_F^2$, with $v_F$ denoting the Fermi velocity. As a
consequence, the density dependence of long-wavelength plasmons in
massless 2D Dirac systems (with $E_{\bf k} = \hbar v_F k$), such
as graphene or borophene, is $\omega_{\rm pl} (q \to 0)\propto
n^{1/4}$. In case of 2D massive Dirac systems with $E_{\bf k} =
\sqrt{\hbar^2 v_F^2 k^2 + \Delta^2}$, with $2\Delta$ denoting the bandgap, 
this density dependence
changes to $\omega_{\rm pl} (q \to 0) \propto n^{1/2}/[n + g
\Delta^2/(4 \pi \hbar^2 v_F^2)]^{1/4}$, where $g$ denotes the spin
and valley degeneracy.\cite{thakur3} Thus, in the limit of low
densities in massive Dirac systems in 2D (more precisely for $n
\ll \Delta^2/(4 \pi \hbar^2 v_F^2)$, we have $\omega_{\rm pl}
\propto n^{1/2}$, as has been shown for the case of monolayer
MoS$_2$, which can be considered as a solid-state example of
massive 2D Dirac system.\cite{R9}

Additionally, plasmon dispersions of n- and p-doped samples in 2D
semiconductors are quite different, due to the marked
electron-hole asymmetry, specifically for higher carrier
concentrations. Consequently, in 2D semiconductors plasmonic
excitations can be tuned by varying the carrier concentration by
means of a back gate.

A distinct advantage of plasmons in 2D semiconductors is the
relatively large and electrically tunable lifetime of plasmons,
which dictates their potential use for technological applications.
Note that, within RPA, the plasmon modes are undamped in the
regions of the $(q,\omega)$ space for which ${\rm
Im}[\Pi(q,\omega)] = 0$, and damping processes (the intraband or
interband single-particle excitations) are activated wherever ${\rm Im}[\Pi(q,\omega)] \neq 0$.
However, in a realistic scenario, several other decay channels of the plasmonic modes exist, 
including scattering from impurities, from phonons, multiparticle excitations etc. In
2D semiconductors with a large band gap, the plasmon damping at
large momenta is due to the excitation of intraband 
single-particle excitations, as in MoS$_2$ \cite{R9} and black phosphorus. 
\cite{R10} By contrast, in gapless systems, such as graphene, or
in small-band-gap materials, like silicene and germanene, plasmons
decay in electron-hole pairs via interband transitions.\cite{R11}

\section{Plasmons in 2D semiconductors}
Dirac plasmons in graphene behave quite differently with respect
to traditional 2D materials, such as III-V semiconductor quantum
wells \cite{PhysRevB.39.12682}, due to the following reasons: (i)
the existence of a pseudospin degree of freedom and (ii) the
relativistic nature of Dirac-cone electrons.

As for the case of graphene, transition-metal dichalcogenides
(MoS$_2$, MoSe$_2$, WTe$_2$ etc.) have a honeycomb lattice without
an inversion center. \cite{R7} The band structure exhibits a
direct gap at the two inequivalent valleys centered at the
high-symmetry points $K$ and $K'= -K$ in the Brillouin zone.\cite{R8} 
As a consequence of time-reversal symmetry, which maps
$k \to -k$ and one valley onto the other, electronic states in a
specific band at $K$ and $K'$ have antiparallel angular momenta.
However, the existence of a finite band gap implies that charge
carriers in 2D semiconductors cannot behave as massless particles,
since they carry a finite effective mass, unlike Dirac fermions in
graphene and topological insulators. For this reason, plasmonics
with 2D semiconductors share features of both graphene and 2DEG
systems.


\begin{figure}[t!]
\centering
\includegraphics[width=0.8\linewidth]{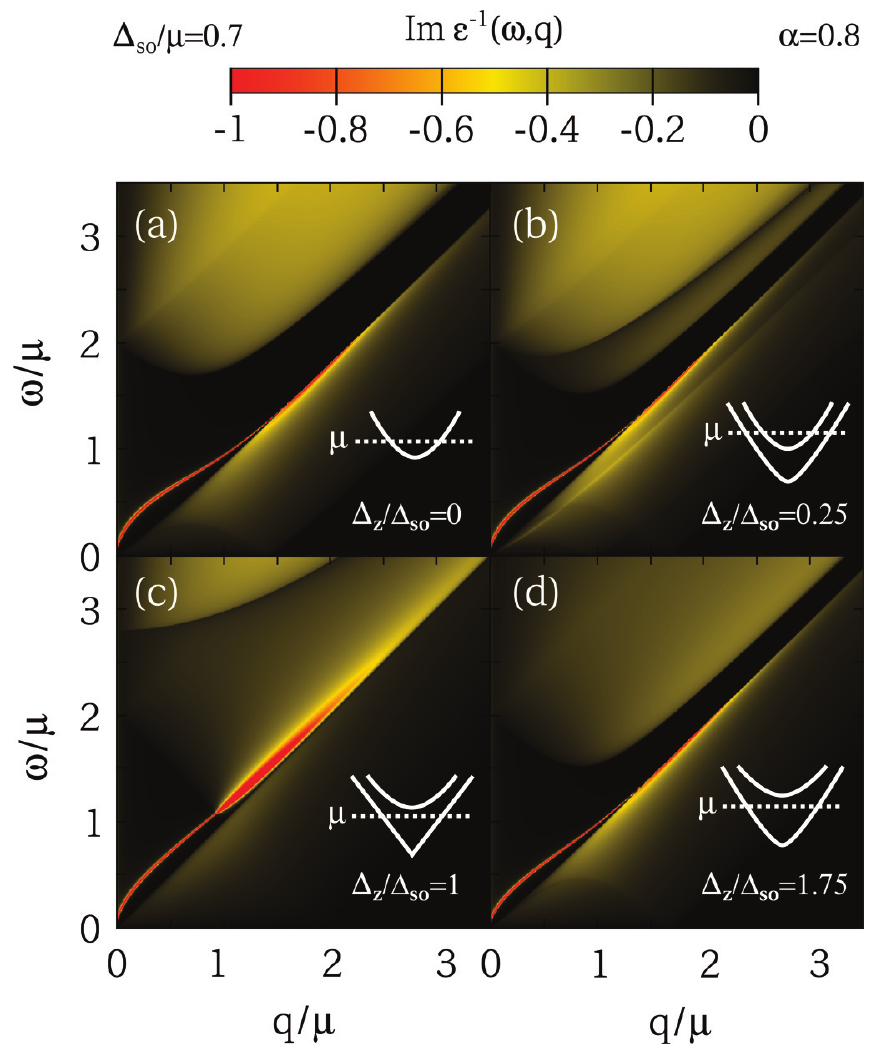}
\caption{The energy loss function for buckled honeycomb lattices,
such as silicene or germanene, along with the plasmon dispersion for a fixed $\Delta_{\rm SO}/\mu = 0.7$
and varying $\Delta_z$. Panel (a) is for $\Delta_z =0$, which is the case of 
a topological insulator, with degenerate spin bands 
as shown in the sketch of the bands. 
(b) is for the case of $\Delta_z/\Delta_{{\rm SO}} <1$, which is the spin-polarized topological-insulator 
phase as indicated in the sketch of the bands.  (c) is for the case of $\Delta_z/\Delta_{\rm SO} =1$, which 
is a valley spin-polarized semimetal with one set of linear band gapless band, and the other massive Dirac like band as shown in the 
sketch of the bands. (d) represents the case of $\Delta_z/\Delta_{{\rm SO}} > 1$, which is an ordinary band insulator. 
The plasmon dispersion is denoted by the red curve in all four panels. 
Here, $2 \pi e^2/\kappa = 0.8$.  In panels (a), (b) and (c) all the band with the lower (higher) energy correspond to the spin up (down) 
bands. Adapted with
permission from Ref.~\cite{R13}.
\label{Fig1}}
\end{figure}


Plasmonics with 2D semiconductors is strongly influenced by the
spin-orbit interaction (SOI), with the subsequent removal of the
spin degeneracy. The coexistence of Bychkov-Rashba (BR) and
Dresselhaus (D) SOI mechanisms induces highly anisotropic
modifications of the static dielectric function \cite{R12} and,
moreover, the SOI also induces a beating of Friedel oscillations,
which can be controlled by external fields. Such a beating
phenomenon has been reported for the cases of MoS$_2$ \cite{R9}
and for buckled honeycomb systems with a remarkable SOI, such as
silicene and germanene \cite{R13}.

The effective low-energy Hamiltonian of buckled honeycomb
structures, such as silicene and germanene, is given by $H = \hbar
v_F (\xi k_x \tau_x + k_y \tau_y) + \Delta_z \tau_z - \xi
\Delta_{\rm SO} \tau_z \sigma_z$. Here, $\sigma$$(\tau$) denotes
the spin (pseudo-spin) degree of freedom and $\xi = \pm 1$ for the
two inequivalent Dirac points $K$ and $K'$. $\Delta_{\rm SO}$
represents the gap induced by SOI (1.55 and 24.0 meV for silicene
and germanene, respectively\cite{R14}). $\Delta_z = E_z \cdot d$
accounts for the A-B sublattice symmetry breaking, where $E_z$ is
the effective external electric field perpendicular to the sample,
including all the screening effects, and $d$ is the perpendicular
distance between the two sublattice planes. The corresponding
low-energy dispersion is given by $E_{\bf k} = \pm \sqrt{(\hbar
v_F |{\bf k}|)^2 + \Delta_{s \xi}^2}$, with the effective gap
being $\Delta_{s \xi} = \Delta_{\rm SO}|s \xi \gamma -1|$. Here,
$s = \pm 1$ denotes the spin, and $\gamma = \Delta_{\rm
SO}/\Delta_{z}$. Interestingly, tuning the vertical electric field
and changing $\Delta_z$ can change the 2D structure from a band
insulator ($\gamma>1$), to a valley spin-polarized metal ($ \gamma
=1$), to a  topological insulator ($\gamma <1$). The most peculiar
characteristics of plasmon modes in buckled honeycomb systems is
that both their screening processes and the plasmon dispersion can
 be tuned by changing $\gamma$.

%
%

The plasmon dispersion in silicene and germanene, when the chemical potential is such that
both the conduction bands are occupied [see inset of Fig.~\ref{Fig1}(b)], is:
\begin{equation}
\omega_{\rm pl} (q) \approx \sqrt{\frac{\alpha_k v_F \mu q}{\hbar} \left(2-\frac{\Delta_+^2 + \Delta_-^2}{\mu^2}\right)},
\end{equation}
where $2 \Delta_{\pm} = \Delta_{\rm SO}|\gamma \mp 1|$, as the
spin-dependent energy gaps, and $\alpha_k$ is the effective fine
structure constant. For the case when only
one band is occupied, or $\Delta_- < \mu < \Delta_{+}$, the
corresponding plasmon dispersion is
\begin{equation}
\omega_{\rm pl} (q) \approx \sqrt{\frac{\alpha_k v_F \mu q}{\hbar} \left(1-\frac{\Delta_-^2}{\mu^2}\right)}.
\end{equation}

We emphasize again that $\omega_{\rm pl} (q) \propto
\sqrt{q}$ is a peculiarity of every 2DEG system with long-range Coulomb interactions.\cite{R6}
By adjusting the electric field, it is possible to tune
the plasmon energy, which undergoes a blue-shift as the ratio
$\gamma$ decreases - see Fig.~\ref{Fig1}. 
Additionally, the damping processes of plasmon modes in
silicene and germanene reflect the change of the topology of the 
energy bands. For $\gamma \to 1$, the small band gap reduces the
lifetime of the plasmon excitation, which could be damped by
thermal fluctuation or disorder scattering.\cite{R15} For $\gamma
\neq 1$, the damping of the plasmon mode is suppressed. By
evaluating $\gamma$, it can be concluded that the lifetime of
plasmonic modes in germanene should be rather higher compared to
silicene.

Wu et. al. \cite{R16} have pointed out the importance of the
interplay of SOI and temperature in silicene-based plasmonics. In
particular, an additional plasmon branch
emerges for $k_B T \approx E_g$, where $k_B$ is the Boltzman constant and $E_g$ is the bandgap. 
Moreover, the lifetime of the plasmon modes is also
affected by temperature. Thus, monitoring the plasmon peak, so as to
probe abrupt changes in its spectral width and dispersion, could
be an useful tool for evaluating temperature-induced changes
in SOI. Using the finite temperature polarization function of
silicene, germanene and other similar buckled honeycomb
structures, it was also shown that the exchange and correlation
energies decrease with increasing temperature.\cite{0953-8984-29-13-135602} 
Besides the low-energy intraband
plasmon, high-energy interband plasmons in such buckled structures
have also been studied, both experimentally
\cite{PhysRevB.97.041401} and theoretically.\cite{PhysRevB.95.085419} 
However, the practical exploitation of
high-energy excitations is generally challenging \cite{JPCC_HEP}.

Among other 2D semiconductors, the anisotropic lattice of black
phosphorus provides novel routes for plasmonics, due to its
subsequent band anisotropy. The plasmon disperses differently due
to the mass anisotropy, where the smaller mass along the
armchair direction leads to higher resonance frequency.\cite{R10} 
For phosphorene, the effective low-energy dispersion in vicinity
of the $\Gamma$ point is given by $H = (u k_y^2 + \Delta) \tau_x +
v_f k_x \tau_y$, where the subscript $x/y$ in $k_{x/y}$ denotes
the direction of the armchair/zigzag edge and $\tau$ denotes the
Pauli matrices. The corresponding band dispersion can be further
approximated by an anisotropic parabolic dispersion, $E_{\bf k} =
E_c + \hbar^2 k_x^2/(2 m_x) + \hbar^2 k_y^2/(2 m_y)$, with $E_c =
\Delta$ being half of the bandgap. The anisotropic band masses are given by
\cite{PhysRevB.96.035422} $m_x = \hbar^2 \Delta/v_f^2  \approx 0.2
m_e$, and $m_y = \hbar^2/(2 u) \approx 1.1 m_e$.

Within the anisotropic mass approximation of the monolayer
phosphorene band structure, the long-wavelength plasmon dispersion
is given by \cite{PhysRevB.96.035422,PhysRevB.91.075422}
\begin{equation} \label{pl1}
\omega_{\rm pl}({\bf q})  = \alpha (\mu - E_c)^{1/2} \left[\cos^2 \theta_q  + \frac{m_x}{m_y} \sin^2\theta_q\right]^{1/2}  \sqrt{q}~,
\end{equation}
where $\theta_q = \tan^{-1}{(q_y/q_x)}$, $\alpha^2 = 2 \pi e^2
g_{\rm 2d}/(m_x \kappa)$, and $g_{\rm 2d} = \sqrt{m_xm_y}/(\pi
\hbar^2 )$ is the 2D density of states for an anisotropic
parabolic band system. Note that the $\sqrt{q}$ dependence of the
plasmon dispersion in all directions for small wave-vectors is
maintained. The anisotropic mass approximation still keeps the
$\omega_{\rm pl} \propto \mu^{1/2}$ dependence.\cite{R10} However, the
anisotropy of long-wavelength ($q \to 0$) plasmon dispersion
$\omega_{\rm pl}(q {\hat x})/\omega_{\rm pl}(q {\hat y}) =
\sqrt{m_y/m_x}$ does not depend on the chemical potential of the
system. For multilayer phosphorene, the band non-parabolicity
caused by interband coupling leads to a plasmon frequency scaling
as $n^{\beta}$ , where $n$ is the carrier concentration, and
$\beta < 1/2$.\cite{R10}
The non-parabolicity effects are generally more prominent for thicker films of phosphorene.

\begin{figure}[t!]
\centering
\includegraphics[width=1\linewidth]{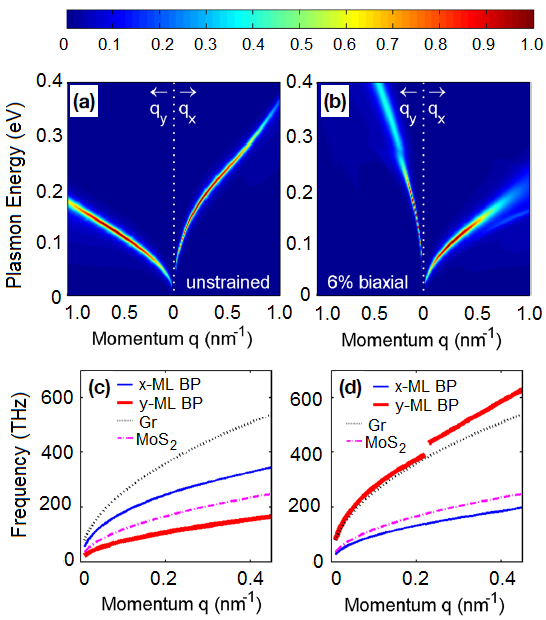}
\caption{ (a) The anisotropic plasmon dispersion of monolayer phosphorene.
The plasmon mass, being larger in the $y$ (zigzag) direction of phosphorene,
leads to reduced plasmon frequencies as $\omega_{\rm pl} \propto
m^{-1/2}$. (b) However, strain inverts the mass anisotropy, and,
consequently, biaxial strained phosphorene has larger plasmon
frequency along the zigzag edge, as compared to the armchair edge.
(c) Comparison of the plasmon dispersion in unstrained graphene
(black dotted curve), MoS$_2$ (magenta dashed-dotted curve) and in
phosphorene along the zigzag edge (blue thin curve) and along the
armchair edge (red thick curve). (d) Same as in (c) but for $6\%$
biaxially strained phosphorene. The plasmon dispersions of
graphene and the MoS$_2$ are for unstrained systems in both
panels (c) and (d). 
The jump in the plasmon dispersion in panel (d) for phosphorene
zigzag edge is due to the presence of a second higher frequency
peak in the loss function. Adapted with permission from
Ref.~\cite{doi:10.1063/1.4914536}.
\label{Fig2}}
\end{figure}

Another useful feature of plasmonics in phosphorene is the
tunability of the band structure via applied strain
\cite{BP_strain1,BP_strain2} or an out-of-plane static electric
field, which can be experimentally implemented via potassium doping.\cite{Kim723} 
This leads to a significant change in the plasmon
dispersion as well \cite{doi:10.1063/1.4914536,R20}. For example, 
strain can invert the anisotropy of the electronic band masses in
phosphorene \cite{BP_strain1}, making $m_x > m_y$, resulting in
higher frequency of the plasmons in along the armchair direction
($\Gamma-X$) compared to the plasmon frequency in the zigzag
direction ($\Gamma-Y$), in contrast to the unstrained case [see
Fig.~\ref{Fig2}]. Based on the tunable anisotropic plasmonic
response, multilayer phosphorene can be used in anisotropic
plasmonic devices.\cite{doi:10.1021/acs.nanolett.5b05166}

In addition, black phosphorus has been shown to behave as a 2D
hyperbolic material.\cite{R17} A hyperbolic material is a highly
anisotropic material, for which the components of the permittivity
parallel ($\epsilon_{\parallel}$) and perpendicular
($\epsilon_{\perp}$) to the crystal axis have opposite sign. 
\cite{R18} The usually elliptic isofrequency curves of the
extraordinary wave become a hyperboloid in black phosphorus. Due
to this unique topology, black phosphorus allows the propagation
of otherwise evanescent waves.

Black phosphorus also exhibits a surface plasmon polariton (SPP)
\cite{R19} with extraordinary tunability. More recently,
femtosecond photo-switching of interface polaritons in black
phosphorus heterostructures has also been demonstrated. 
\cite{Fphotoswitching} The hybrid phonon-plasmon-polariton mode
is transient, switchable, exhibits large propagation length, and
is likely to be very useful in polariton-based mid-infrared
optoelectronic devices.\cite{Fphotoswitching} 

The plasmonic spectrum of bilayer phosphorene is characterized by two plasmon modes \cite{R20}, representing in-phase and out-of-phase oscillations of the carrier density in the two planes, with a dispersion $\omega_{+}(q) \propto \sqrt{q}$ and
$\omega_-(q) \propto q$, respectively.\cite{171007808S} Disorder induces the loss of the decoherence of the $\omega_{-}(q)$ mode.

An electric field applied perpendicular to bilayer phosphorene can be used to tune the dispersion of the plasmon modes.\cite{R20} For sufficiently large electric field, bilayer phosphorene enters in a topological phase with Dirac-like crossing of bands in one direction and gapped in the other direction. Consequently, the excitation spectrum 
has different features in this limit, with highly coherent plasmon modes, which are gapped in the armchair direction.
Interestingly, the strength of screening of electric fields in black phosphorus \cite{R10} ranges between the strong coupling regime characteristic of graphene \cite{R21} and the reduced screening properties of transition-metal dichalcogenides.\cite{R22}

More recently, several monolayer polymorphs of boron have been experimentally demonstrated. \cite{Borophene,Mannix1513} In particular, $8-Pmmn$ borophene polymorph has been shown to host massless Dirac fermions with anisotropic and tilted Dirac cone \cite{PhysRevLett.112.085502,PhysRevLett.118.096401}. This leads to anisotropic plasmon dispersion, along with anisotropic static screening and Friedel oscillations.\cite{PhysRevB.96.035410}

\begin{figure*}[t!]
\centering
\includegraphics[width=0.8\linewidth]{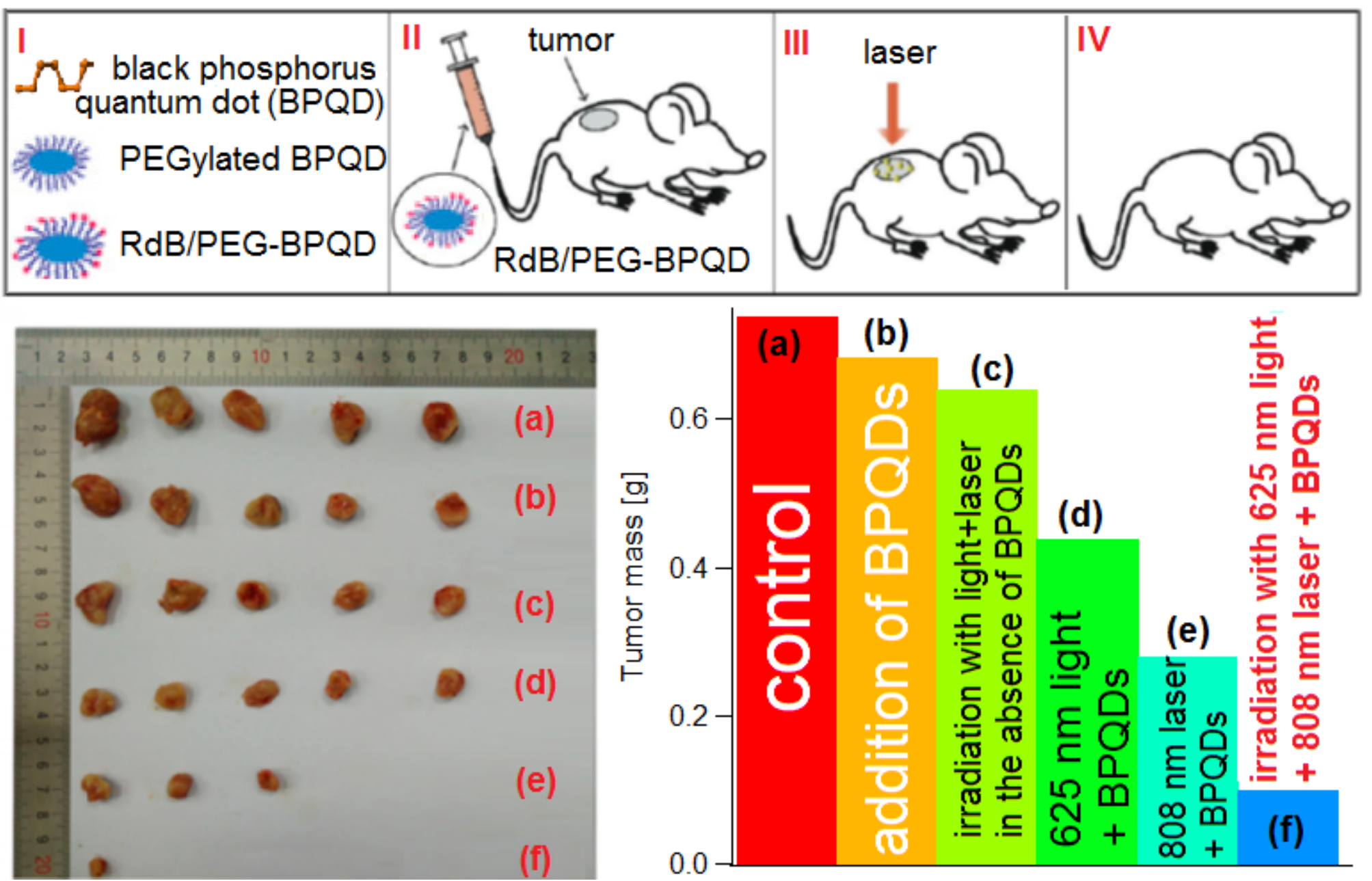}
\caption{In the top part, panel I shows a black phosphorus quantum
dot (BPQD), a PEGylated BPQD (where PEG stands for polyethylene
glycol), and finally a RdB/PEG BPQD (where RdB stands for
rhodamine B). In panel II, the injection of solution containing
RdB/PEG BPQDs in a mouse bearing a tumor mass is depicted. Panels
III and IV represent the photothermal treatment and the results,
i.e. the disappearance of the tumor mass. In the bottom part,
(left) photographs of tumors collected from different groups of
mice at the end of treatments (16 days) and (right) the average
weight of tumors collected from different groups of mice. (group
a: control; group b: BPQDs; group c: irradiation with 625 nm light + 808 nm
laser; group d: 625 nm light in the presence of BPQDs; group e: 808 nm laser in the presence of
BPQDs; group f: 625 nm light + 808 nm laser in the presence of BPQDs). Adapted
with permission from Ref.~\cite{Figmice}.
\label{Fig3}}
\end{figure*}

\section{Potential Applications}
The monolayer thickness of 2D semiconductors is an important
advantage in many applications, such as field-effect transistors
(FETs) for high-performance electronics, sensing applications, and
flexible electronics. Nevertheless, in the fields of
optoelectronics and photonics the monolayer thickness represents a
major challenge concerning the interaction with light, with
usually insufficient light emission and absorption. However,
several researchers have reported an enhancement of
photoluminescence from large-area monolayer MoS$_2$ using
plasmonic noble-metal (Ag or Au) nanostructures.\cite{R23,R24,R25} 
As a matter of fact, light can be trapped at
MoS$_2$ with field enhancement near the plasmonic nanostructures
\cite{R26}, with potential exploitation for innovative
optoelectronic devices, such as photodetectors and emitters.
However, it should be considered that the localized temperature
increase at the irradiated nanoparticles, due to photothermal
effects, also implies a structural phase transition in MoS$_2$. 
\cite{R27}

Studies of photocurrent generation at MoS$_2$-metal junctions
\cite{R28} have indicated that the polarized photocurrent response
can be interpreted in terms of the polarized absorption of light
by the plasmonic metal electrode, its conversion into hot
electron-hole pairs, and subsequent injection into MoS$_2$. These
fundamental studies shed light on the knowledge of photocurrent
generation mechanisms in metal-semiconductor junctions, opening
the door for engineering future optoelectronics through SPP in 
2D materials.

To extend the operational range of plasmonics with 2D materials,
their chemical modification via intercalation is a suitable route.
As an example, the electrochemical intercalation on lithium in
MoS$_2$ flakes induces the emergence of plasmon resonances in the
visible and near-UV range of the electromagnetic spectrum. 
\cite{R29}

As for conventional plasmonics, biosensing applications have been explored also by using 2D semiconductors.\cite{R29,R30}

Recent findings for black phosphorus \cite{R31,R32},
MoS$_2$\cite{R33} and MoSe$_2$ \cite{R34} indicate that 2D
semiconductors can be suitable also for thermoplasmonics
\cite{R35,R36,R37,R38}, i.e., the thermal heating associated to
optically resonant plasmonic excitations in nanoparticles.
Thermoplasmonics relies on the control of nanoscale thermal
hotspots by light irradiation.\cite{R39} Photothermal effects may
be used for enabling plasmon-induced reactions \cite{R40,R41,R42}
or for photothermal treatment of epithelial cancer.\cite{R43,R44,R45,R46} 
As sketched in Figure 3, it has been shown
that tumor-bearing mice were entirely convalesced after
photothermal treatment with nanoparticles of 2D semiconductors.\cite{R32} 
As a matter of fact, quantum dots of 2D semiconductors
are characterized by an excellent near-infrared (NIR) photothermal
conversion efficiency (28.4 \% for black phosphorus \cite{R31} and
46.5\% for MoSe$_2$ \cite{R34}), a large extinction coefficient,
as well as good photostability and enhanced stability in
physiological medium. In vitro experiments demonstrate that 2D
semiconductors have negligible cytotoxicity and can efficiently
kill cancer cells under laser irradiation in the visible and NIR
range of the electromagnetic spectrum (Figure 3), ensuring the
feasibility of thermoplasmonic cancer treatment with 2D
semiconductors.

\begin{figure*}[t!]
\centering
\includegraphics[width=0.8\linewidth]{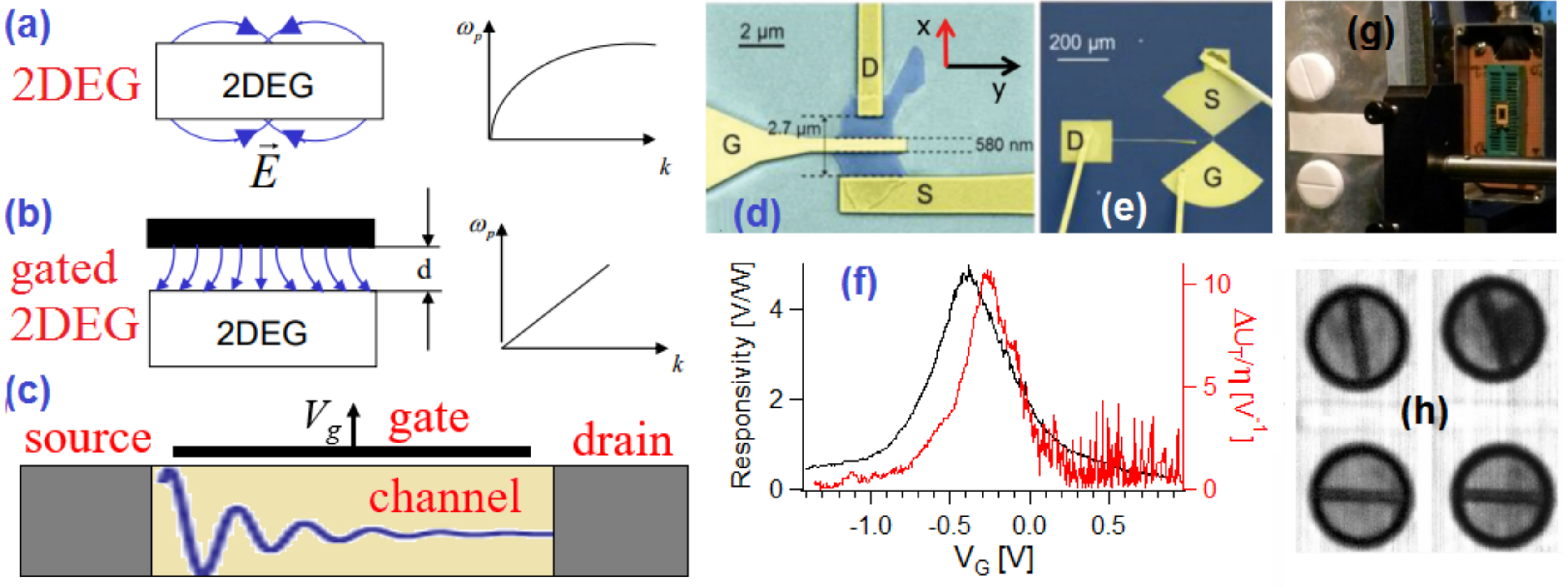}
\caption{Panels (a) and (b) display the dispersion relation of the
plasmon frequency in 2DEG and gated 2DEG, respectively. Overdamped
propagation of plasma waves, inevitable at room temperature, is
sketched in panel (c). In panels (d) and (e) the nanofabrication
of THz photodetectors is depicted. Panel (f) shows a comparison 
between the experimental photoresponse (black curve) and the calculated 
photoresponse within the framework of the Dyakonov-Shur model\cite{R48} (red curve). 
The two curves qualitatively match, even if the agreement is not
perfect, thus implying that the THz photodetection mechanism is
not uniquely that one connected with plasma waves. Concerning
large-area transmission THz imaging experiments, in panel (g) two
tablets before and after injecting a drop of water are displayed.
The water content is clearly imaged in the THz transmission image
in panel (h). Adapted with permission from
Ref.~\cite{R51} \label{Fig4}}
\end{figure*}

The occurrence of plasmonic modes in the Terahertz (THz) range in
the whole class of 2D semiconductors makes them the ideal
candidates for plasma-wave THz photodetection. THz represents one
of the more exciting technological challenges with extraordinary
prospect in the fields of wireless communications, homeland
security, night-vision, gas sensing and biomedical applications.\cite{0022-3727-50-4-043001}
The
ability to convert light into an electrical signal with high
efficiencies and controllable dynamics is a major need in
photonics and optoelectronics.\cite{R47} The active channel of a
FET hosting a 2DEG acts as a cavity for plasma waves, which are
described within a hydrodynamic model developed by Dyakonov and Shur.\cite{R48} 
Note that plasma waves propagating in the FET
channel cannot be simply identified with the well-known plasmons
of a 2DEG \cite{R50}, because of the presence of the gate of the
FET which induces a linear dispersion relation of the plasma waves
(Figure 4, panels a-b). The generated photoresponse $\Delta u_T$
can be deduced from the transfer characteristics of the FET via
the relation \cite{R53,R54}:
\begin{equation}
\Delta u_T = \frac{1}{\sigma} \cdot \frac{d\sigma}{dV_G} \cdot
\left( \frac{R_L}{R_L + \sigma^{-1}}\right)\cdot \eta,
\end{equation}
where $\sigma$ is the channel conductivity, $R_L$ is the finite
impedance of the measurement setup including the readout
circuitry, and the constant $\eta$ represents the
antenna-dependent coupling efficiency.

In general, the resonant detection, due to the rectification
induced by plasma waves, is observed only at cryogenic
temperatures. At room temperature, the plasma-wave oscillations
are overdamped (Figure 4c), but the rectification mechanism is
still efficient and enables a broadband THz detection and imaging
\cite{R49} (see Figure 4, panels g and h for the case of
black-phosphorus THz photodetectors). Recently, Viti et. al.
\cite{R51,R52} used the plasma-wave excitation in the active
channel of a phosphorene-based FET (Figure 4, panels c-d) for
devising a THz photodetector, whose capabilities are demonstrated
by imaging of macroscopic samples, in real time and in a realistic
setting (Figure 4, panels g and h). The obtained value of the
noise-equivalent power indicates higher performance of 2D
semiconductors compared with Dirac materials, such as graphene
\cite{R53} or topological insulators.\cite{R54}

Another promising prospect of plasmonics with 2D semiconductors is
related to the fabrication of van der Waals heterostructures
\cite{R55}, i.e. 2D materials stacked with hexagonal boron
nitride. As a matter of fact, recently, graphene stacked with
hexagonal boron nitride (h-BN) has been found to act as a
hyperbolic metamaterial \cite{R56} with unprecedently high plasmon
lifetime and high field confinement.\cite{R57} Tunability is
originated from the hybridization of SPP of the 2D material with
hyperbolic phonon polaritons in h-BN \cite{R58}, so that the
eigenmodes of the van der Waals heterostructure are hyperbolic
plasmon-phonon polaritons.

\section{Conclusions}
2D materials beyond graphene offer very interesting plasmonic properties, which are of great interest both from a 
fundamental as well as technological perspective. A great advantage of 2D materials is the longer lifetime and the tunability of both plasmon dispersion and damping by changing the doping, intercalating chemical species, applying vertical electric fields etc. This facilitates their interaction with light, leading to 
localization and guiding of light into electrical signals, which 
can be technologically used for devising plasmonic devices for diverse applications, ranging from nanoelectronics, nanophotonics and nanomedicine.

\section*{Acknowledgement}
MSV acknowledge financial support of the European Union ERC Consolidator Grant SPRINT (681379). 
\bibliography{MR_refs} 
\bibliographystyle{rsc} 
\end{document}